# Version 7
# Structured psychosocial stress and therapeutic failure

Rodrick Wallace, Ph.D.
The New York State Psychiatric Institute
Deborah Wallace, Ph.D.
Joseph L. Mailman School of Public Health
Columbia University
*

October 24, 2003

## Abstract

Generalized language-of-thought arguments appropriate to interacting cognitive modules permit exploration of how disease states interact with medical treatment. The interpenetrating feedback between treatment and response creates a kind of idiotypic hall of mirrors generating a synergistic pattern of efficacy, treatment failure, adverse reactions, and patient noncompliance which, from a Rate Distortion perspective, embodies a distorted image of externally-imposed structured psychosocial stress. For the US, accelerating spatial and social diffusion of such stress enmeshes both dominant and subordinate populations in a linked system which will express itself, not only in an increasingly unhealthy society, but in the diffusion of therapeutic failure, including, but not limited to, drug-based treatments.

## I. Introduction

Antibiotic therapies are under siege from evolutionary adaptation by pathogens. Malignancies routinely evolve out from under anti-cancer drugs. Antiretroviral therapies against human immunosuppressive virus (HIV), although often able to prolong life, ultimately fail for much the same reason, and multiple-drug-resistant HIV, (MDR-HIV), like MDR-tuberculosis, is an increasing threat. The role of social processes in some of these matters is well understood: for example, public policies of 'planned shrinkage' in New York City (Wallace and Wallace, 1998) created such degrees of social disintegration that many individual courses of drug treatment were repeatedly interrupted, generating MDR strains of TB whose address required an inordinately expensive program of highly labor-intensive directly observed therapy (DOT) estimated to cost $40,000 per patient (Wallace and Wallace, 1997; D. Wallace, 1994).

*Address correspondence to R. Wallace, PISCS Inc., 549 W. 123 St., Suite 16F, New York, NY, 10027. Telephone (212) 865-4766, rdwall@ix.netcom.com. Affiliations are for identification only.

What is less understood is that the diffusing structured psychosocial stress responsible for the unhealthy societies described by Wilkinson (1996) appears able to affect other therapeutic agents and agencies besides antibiotic drugs. Here we will, in particular, outline how other drug therapies are likely to come under siege, especially in the United States. From a public health perspective this is of limited significance, since salubrious living and working conditions, and egalitarian social relations, rather than effective drugs, are the principal determinants of population health. That being said, spreading decline of drug efficacy and rising rates of adverse reactions and patient noncompliance would nonetheless constitute a significant exacerbation of disease burden.

From the perspective of the economics of the pharmaceutical industry, however, the avalanche of drug failure, side effects, and noncompliance we predict for asthma, diabetes, obesity, hypertension, coronary heart disease, hormonal cancers, depression and other mental disorders, is likely to prove devastating. The extraordinary difficulty of bringing new drugs to market will, if we are correct, soon be markedly compounded. Indeed, recent work by Lazarou et al. (1998) suggests that, even at present, adverse drug reactions are already the fourth commonest cause of death in the US.

Adverse drug reactions, to paraphrase Pirmohamed et al. (2002), are typically either consistent with the known pharmacology of the drug, representing an augmentation of its known effects, and are dose-dependent, or else are bizarre responses to idiosyncratic induced hypersensitivity, with highly variable outcomes depending on both the drug and the patient.

The essential role of stress in drug efficacy was recognized some time ago by Downing and Rickels (1982), work rediscovered recently by Haller and colleagues (e.g. Haller and Halasz, 2000; Haller, 2001). Downing and Rickels (1982) found that the anxiolytic efficacy of chlordiazepoxide and diazepam was markedly reduced in patients experiencing unfavorable life events during drug treatment, compared with both patients experiencing favorable life events or no major events. As Haller (2001) put it,

> "To our knowledge, the impact of this finding [i.e. Downing and Rickels, (1982)] was relatively small, and animal research on the topic was not prompted by it... [nonetheless] these and other similar find-ings show that drug efficacy is not constant, and [epi]genetic factors (e.g. stress) have a strong modulator effect."

Patient medication noncompliance, to paraphrase the conventional perspective taken by Miura et al. (2001), is already a serious factor limiting the effectiveness of medical treatments. For instance, the classic study of Sackett and Haynes (1976), found that, even under the best of circumstances, over 30 percent of patients skipped their prescribed doses regardless of their disease, prognosis, or symptoms.

Many such problems are related to maintaining long-term therapy in patients with chronic disease such as hypertension. Factors encouraging noncompliance in long-term therapy include the cost of medication, lack of written instructions, nonparticipation of the patient in designing the treatment plan, lack of patient education about disease, side effects, and inconvenient dosing schedules. These factors may enhance the frequency of patient noncompliance as the duration of drug therapy is prolonged.

Langer (1999) takes a view more in concert with current medical anthropology:

> "Care providers [have socially constructed] behavior as compliant when it was patterned by their expectations and noncompliant when behavior deviated from their [culturally and socially conditioned] expectations... [Their] [l]ack of awareness of cultural issues increases [patient-provider] social distance, breaks down communications, and precipitates misconceptions between minority patients and their health care providers. Therefore, opportunities for patient dissatisfaction and noncompliance increase...
> Brock and Salinsky (1993) use the term 'therapeutic alliance' to denote a process in which the health care provider communicates an assessment of the patient's problem and coordinates a practical management plan that is conducive to patient compliance. This assessment must consider and integrate information about all the systems in which the person exists: biological, psychological, informal/formal social support system, and cultural... Elevating the patient's status within the therapeutic alliance increases the likelihood of... participation, that is, enhanced compliance."

Here we will suggest that stress itself is not undifferentiated, like pressure under water, but often has a complicated grammar and syntax which can, in a sense, carve a distorted image of that structure onto basic human physiological and psychological structures and their responses to medical intervention, including but not limited to, pharmaceuticals.

We will, quite formally, synthesize these matters at and across physiological, psychological, and psychosocial levels of organization, particularly considering the impact of structured stress on the therapeutic alliance.

Essential clues are to be found in observed patterns of population-level disparity in drug efficacy, adverse reactions, and compliance across ethnic groups, once we have discounted contemporary pharmacogenetic ideology.

According to Burroughs et al. (2002), pharmacogenetic research over the past few decades has uncovered significant differences among racial and ethnic groups in the metabolism, clinical effectiveness, and side-effect profiles of many clinically important drugs, a matter which, they claim, can be attributed to "...genetic factors that underlie varying responses to medicines observed among different genetic and racial groups".

An unsigned editorial commentary in a recent issue of *Nature:genetics* (vol. 29, no.3, 239, 2001) is somewhat more guarded, but reaches a similar conclusion, focused, however, at the individual rather than the group locus-of-control:

> "Adverse drug reactions or failure to respond to certain drugs can be influenced by polymorphisms in genes responsible for metabolizing drugs. [While] [t]he frequency of such polymorphisms has been found to vary between populations of common ancestry... a number of studies... have shown that average differences in the genetic determinants of drug response... [between] groups... are relatively small and there is considerable overlap between groups. *Once all the genes that contribute to drug response are identified, doctors will be able to prescribe drugs based on patients' [individual] genotypes... this is the promise of the Human Genome Project.*" (emphasis added)

Depending on mainstream perspective, then, it is 'genetic polymorphisms' at either the group or individual level which are responsible for most of the variance in patient response to drug therapy.

Currently one of the most discussed examples is the finding that, while 49 percent of white male patients benefited from an angiotensis-converting enzyme inhibitor for heart failure, only 14 percent of African-American males did so (Exner et al., 2001). The point of debate within the biomedical mainstream is not whether this result is a actually attributable to genetic polymorphism or not, but, rather, appears limited to questions regarding the level of organization at which such polymorphism should be investigated, i.e. the racial/ethnic group or the individual.

The reader may have noted the scientific mystery implicit in these results: since the distribution of polymorphisms related to drug efficacy appears to greatly overlap between 'racial' subgroups precisely as the *Nature:genetics* editorial commentary states, the three-fold difference in such efficacy between white and African-American males cannot be entirely, or indeed, even significantly, explained by those polymorphisms.

What is really going on?

Here we raise a larger and far more fundamental locus-of-control question about therapeutic efficacy, compliance, and adverse reactions to therapy, including, but not limited to, drugs. That locus of control is the embedding pattern of structured psychosocial stress exemplified by figure 1, which redisplays material from Singh-Manoux (2003). It shows, for men and women separately, self-reported health as a function of self-reported status rank, where 1 is high and 10 low rank, among some 7000 male and 3400 female London-based office staff, aged 35-55 working in 20 Civil Service departments during the late 1990's. Self-reported health is a highly significant predictor of future morbidity and mortality. The results for both sexes are virtually indistinguishable in what is a kind of toxicological dose-response curve, showing physiological response against a 'dosage' of hierarchy that may include both measures of highly structured stress and real availability of resources (Link and Phelan, 2000). Note that low status staff approach the critical 'ED-50' stage at which half the population shows impact from the dosage.

Figure 1 and its generalizations have been the subject of a long scientific debate regarding the way in which the health of those on the lower left hand side are linked to those on the upper right through various mechanisms. Wilkinson, (1996), for example, describes how the rich of nations with lesser disparities between rich and poor are markedly healthier than the rich in nations – like the US – having greater disparities.

In precisely this regard, figure 2 shows the Black vs. White diabetes death rates (per 100,000) covering the period 1979-97 for the US, and figure 3 the corresponding hypertension death rates over the same time span. 1997 is, in both cases, the point at the upper right. While the Black rates are elevated compared to the White in both cases, the relentless increases are highly correlated indeed, suggesting a single underlying cause which enmeshes both dominant and subordinate populations.

Figure 4 suggests etiology. It shows the rapidly increasing percent of total US income accounted for by the highest five percent of earners as a function of the integral of the number of manufacturing jobs lost since 1980. This latter is an index of permanently dispersed social, economic, and political capital for vast sectors of the population. While US deindustrialization has most profoundly affected minority urban neighborhoods (e.g. Massey, 1990), it has also devastated many white working-class communities (e.g. Pappas, 1989). See Wallace et al. (1999) or Wallace and Wallace (2003) for a more complete discussion.

The general inference is that deindustrialization and its associated phenomena have exacerbated a pathogenic social hierarchy enmeshing both dominant and subordinate populations into trajectories of serious developmental disorder. Absent large-scale social, economic, and political reforms, this process will certainly continue.

We intend to generalize this argument to the role of structured psychosocial stress in therapeutic efficacy, failure, adverse reactions, and patient noncompliance.

Within the US, the particular analogs to figure 1 inevitably involve the aftermath of a slave economy that persisted into the middle 19th Century and has smoothly evolved into the current, and newly-resurgent, system of American Apartheid (e.g. Massey and Denton, 1993). The latter is inextricably convoluted with a market economy which, in the aftermath of deindustrialization and the decline of unions, gives few rights, little stability, and sharply declining real resources for the vast majority of its participants. In addition, deliberate policies of ethnic cleansing ranging from urban renewal to planned shrinkage have left virtually all urban African-American neighborhoods looking like Dresden after the firebombing (e.g. Wallace and Wallace, 1998), with the inevitable individual and community-scale horrors consequent on such acts.

It is little wonder that African-American males do not respond well to certain classes of drugs. Given the implications of figures 2-4, it seems unlikely such failure will be restricted to them, however.

We will use a mathematical modeling strategy to investigate more precisely how structured psychosocial stress, like that of figure 1, might insinuate itself into physiological and other response to medical therapy, including but not limited to drug efficacy and adverse drug reactions.

The results of this exercise suggest that questions regarding cross sectional and longitudinal structured stress at the individual level might, in the vast majority of cases, provide more useful information in designing and monitoring medical interventions, including drug therapies, than individualized genetic profiles. This conclusion is, however, modulated by the inescapably irreducible consequences of such stress at the population level, a matter to which we will repeatedly return, and which is profoundly implicated in our predictions of spreading drug failure.

We begin by examining mind/body interaction as a composite structure of within-individual 'cognitive modules' enmeshed with an immediate 'sociocultural network', all embedded in a larger context of structured psychosocial stress. We thus implicitly embrace 'comorbidity' between chronic mental and physical disorder, and examine therapeutic intervention in such a structure.

This is no small matter, and leads to use of cutting-edge methods. Indeed, further comment on our methodology is appropriate:

Here we adapt recent advances in understanding punctuated equilibrium in evolutionary process (e.g. Wallace, 2002b; Wallace and Wallace, 1998, 1999) to the question of how embedding structured psychosocial stress affects the interaction of mind, body, pathology, and medical treatment. We specifically seek to determine how the synergism of stress, cognitive submodules, and therapeutic intervention, might be constrained by certain of the asymptotic limit theorems of probability.

We know that, regardless of the probability distribution of a particular stochastic variate, the Central Limit Theorem ensures that long sums of independent realizations of that variate will follow a Normal distribution. Analogous constraints exist on the behavior of information sources – both independent and interacting – and these are described by the limit theorems of information theory. Imposition of phase transition formalism from statistical physics, in the spirit of the Large Deviations Program of applied probability, permits concise and unified description of evolutionary and cognitive 'learning plateaus' which, in the evolutionary case, are interpreted as evolutionary punctuation (e.g. Wallace, 2002a, b). This approach provides a 'natural' means of exploring punctuated processes in the effects of structured stress on mind-body interaction in the context of therapeutic intervention.

The model, as in the relation of the Central Limit Theorem to parametric statistical inference, is almost independent of the detailed structure of the interacting information sources inevitably associated with cognitive process, important as such structure may be in other contexts. This finesses some of the profound ambiguities associated with 'dynamic systems theory' and 'deterministic chaos' treatments in which the existence of 'dynamic attractors' depends on very specific kinds of differential equation models akin to those used to de-scribe ecological population dynamics, chemical processes, or physical systems of weights-on-springs. Cognitive phenomena are neither well-stirred Erlenmeyer flasks of reacting agents, nor distorted mechanical clocks, and the application of nonlinear dynamic systems theory to cognition will likely be found to involve little more than hopeful metaphor. Indeed, much of contemporary nonlinear dynamics can be subsumed within our formalism through symbolic dynamics discretization techniques (e.g. Beck and Schlogl, 1995).

In contrast, it seems actually possible to uncover the grammar and syntax of structured psychosocial stress and the function of cognitive submodules, and to express their relations in terms of empirically observed regression models relating measurable biomarkers, behaviors, beliefs, feelings, and so on. Our analysis will focus on the eigenstructure of those models, constrained by the ergodic and other properties of information sources.

Clearly our approach takes much from parametric statistics, and, while idiosyncratic 'nonparametric' models may be required in special cases, we may well capture the essence of the most common relevant phenomena.

### Some cognitive modules of human biology

**1. Immune function** Atlan and Cohen (1998) have proposed an information-theoretic cognitive model of immune function and process, a paradigm incorporating cognitive pattern recognition-and-response behaviors analogous to those of the central nervous system. This work follows in a very long tradition of speculation on the cognitive properties of the immune system (e.g. Tauber, 1998; Podolsky and Tauber, 1998; Grossman, 1989, 1992, 1993a, b, 2000).

From the Atlan/Cohen perspective, the meaning of an antigen can be reduced to the type of response the antigen generates. That is, the meaning of an antigen is functionally defined by the response of the immune system. The meaning of an antigen to the system is discernible in the type of immune response produced, not merely whether or not the antigen is perceived by the receptor repertoire. Because the meaning is defined by the type of response there is indeed a response repertoire and not only a receptor repertoire.

To account for immune interpretation Cohen (1992, 2000) has reformulated the cognitive paradigm for the immune system. The immune system can respond to a given antigen in various ways, it has 'options.' Thus the particular response we observe is the outcome of internal processes of weighing and integrating information about the antigen.

In contrast to Burnet's view of the immune response as a simple reflex, it is seen to exercise cognition by the interpolation of a level of information processing between the antigen stimulus and the immune response. A cognitive immune system organizes the information borne by the antigen stimulus within a given context and creates a format suitable for internal processing; the antigen and its context are transcribed internally into the 'chemical language' of the immune system.

The cognitive paradigm suggests a language metaphor to describe immune communication by a string of chemical signals. This metaphor is apt because the human and immune languages can be seen to manifest several similarities such as syntax and abstraction. Syntax, for example, enhances both linguistic and immune meaning.

Although individual words and even letters can have their own meanings, an unconnected subject or an unconnected predicate will tend to mean less than does the sentence generated by their connection.

The immune system creates a 'language' by linking two ontogenetically different classes of molecules in a syntactical fashion. One class of molecules are the T and B cell receptors for antigens. These molecules are not inherited, but are somatically generated in each individual. The other class of molecules responsible for internal information processing is encoded in the individual's germline.

Meaning, the chosen type of immune response, is the outcome of the concrete connection between the antigen subject and the germline predicate signals.

The transcription of the antigens into processed peptides embedded in a context of germline ancillary signals constitutes the functional language of the immune system. Despite the logic of clonal selection, the immune system does not respond to antigens as they are, but to abstractions of antigens-in-context.

**2. Tumor control** We propose that the next cognitive submodule after the immune system is a tumor control mechanism which may include immune surveillance, but clearly transcends it. Nunney (1999) has explored cancer occurrence as a function of animal size, suggesting that in larger animals, whose lifespan grows as about the 4/10 power of their cell count, prevention of cancer in rapidly proliferating tissues becomes more difficult in proportion to size. Cancer control requires the development of additional mechanisms and systems to address tumorigenesis as body size increases – a synergistic effect of cell number and organism longevity. Nunney concludes

> "This pattern may represent a real barrier to the evolution of large, long-lived animals and predicts that those that do evolve ... have recruited additional controls [over those of smaller animals] to prevent cancer."

In particular, different tissues may have evolved markedly different tumor control strategies. All of these, however, are likely to be energetically expensive, permeated with different complex signaling strategies, and subject to a multiplicity of reactions to signals, including those related to psychosocial stress. Forlenza and Baum (2000) explore the effects of stress on the full spectrum of tumor control, ranging from DNA damage and control, to apoptosis, immune surveillance, and mutation rate. Elsewhere (R. Wallace et al., 2003) we argue that this elaborate tumor control strategy, at least in large animals, must be at least as cognitive as the immune system itself, which is one of its components: some comparison must be made with an internal picture of a healthy cell, and a choice made as to response: none, attempt DNA repair, trigger programmed cell death, engage in full-blown immune attack. This is, from the Atlan/Cohen perspective, the essence of cognition.

**3. The HPA axis** The hypothalamic-pituitary-adrenal (HPA) axis, the flight-or-fight system, is clearly cognitive in the Atlan/Cohen sense. Upon recognition of a new perturbation in the surrounding environment, memory and brain or emotional cognition evaluate and choose from several possible responses: no action needed, flight, fight, helplessness (i.e. flight or fight needed, but not possible). Upon appropriate conditioning, the HPA axis is able to accelerate the decision process, much as the immune system has a more efficient response to second pathogenic challenge once the initial infection has become encoded in immune memory. Certainly hyperreactivity in the context of post-traumatic stress disorder (PTSD) is a well known example. Chronic HPA axis activation is deeply implicated in visceral obesity leading to diabetes and heart disease, via the leptin/cortisol diurnal cycle (e.g. Bjorntorp, 2001).

**4. Blood pressure regulation** Rau and Elbert (2001) review much of the literature on blood pressure regulation, particularly the interaction between baroreceptor activation and central nervous function. We paraphrase something of their analysis. The essential point, of course, is that unregulated blood pressure would be quickly fatal in any animal with a circulatory system, a matter as physiologically fundamental as tumor control. Much work over the years has elucidated some of the mechanisms involved: increase in arterial blood pressure stimulates the arterial baroreceptors which in turn elicit the baroreceptor reflex, causing a reduction in cardiac output and in peripheral resistance, returning pressure to its original level. The reflex, however, is not actually this simple: it may be inhibited through peripheral processes, for example under conditions of high metabolic demand. In addition, higher brain structures modulate this reflex arc, for instance when threat is detected and fight or flight responses are being prepared. This suggests, then, that blood pressure control cannot be a simple reflex, but is, rather, a broad and actively cognitive modular system which compares a set of incoming signals with an internal reference configuration, and then chooses an appropriate physiological level of blood pressure from a large repertory of possible levels, i.e. a cognitive process in the Atlan/Cohen sense. The baroreceptors and the baroreceptor reflex are, from this perspective, only one set of a complex array of components making up a larger and more comprehensive cognitive blood pressure regulatory module.

**5. Emotion** Thayer and Lane (2000) summarize the case for what can be described as a cognitive emotional process. Emotions, in their view, are an integrative index of individual adjustment to changing environmental demands, an organismal response to an environmental event that allows rapid mobilization of multiple subsystems. Emotions are the moment-to-moment output of a continuous sequence of behavior, organized around biologically important functions. These 'lawful' sequences have been termed behavioral systems by Timberlake (1994).

Emotions are self-regulatory responses that allow the efficient coordination of the organism for goal-directed behavior. Specific emotions imply specific eliciting stimuli, specific action tendencies including selective attention to relevant stimuli, and specific reinforcers. When the system works properly, it allows for flexible adaptation of the organism to changing environmental demands, so that an emotional response represents a *selection* of an appropriate response and the inhibition of other less appropriate responses from a more or less broad behavioral repertoire of possible responses. Such choice, we will show, leads directly to something closely analogous to the Atlan and Cohen language metaphor.

Damasio (1998) concludes that emotion is the most complex expression of homeostatic regulatory systems. The results of emotion serve the purpose of survival even in nonminded organisms, operating along dimensions of approach or aversion, of appetition or withdrawal. Emotions protect the subject organism by avoiding predators or scaring them away, or by leading the organism to food and sex. Emotions often operate as a basic mechanism for making decisions without the labors of reason, that is, without resorting to deliberated considerations of facts, options, outcomes, and rules of logic. In humans learning can pair emotion with facts which describe the premises of a situation, the option taken relative to solving the problems inherent in a situation, and perhaps most importantly, the outcomes of choosing a certain option, both immediately and in the future. The pairing of emotion and fact remains in memory in such a way that when the facts are considered in deliberate reasoning when a similar situation is revisited, the paired emotion or some aspect of it can be reactivated. The recall, according to Damasio, allows emotion to exert its pairwise qualification effect, either as a conscious signal or as nonconscious bias, or both, In both types of action the emotions and the machinery underlying them play an important regulatory role in the life of the organism. This higher order role for emotion is still related to the needs of survival, albeit less apparently.

Thayer and Friedman (2002) argue, from a dynamic systems perspective, that failure of what they term inhibitory processes which, among other things, direct emotional responses to environmental signals, is an important aspect of psychological and other disorder. Sensitization and inhibition, they claim, sculpt the behavior of an organism to meet changing environmental demands. When these inhibitory processes are dysfunctional – choice fails – pathology appears at numerous levels of system function, from the cellular to the cognitive.

Thayer and Lane (2000) also take a dynamic systems perspective on emotion and behavioral subsystems which, in the service of goal-directed behavior and in the context of a behavioral system, they see organized into coordinated assemblages that can be described by a small number of control parameters, like the factors of factor analysis, revealing the latent structure among a set of questionnaire items thereby reducing or mapping the higher dimensional item space into a lower dimensional factor space. In their view, emotions may represent preferred configurations in a larger 'state-space' of a possible behavioral repertoire of the organism. From their perspective, disorders of affect represent a condition in which the individual is unable to select the appropriate response, or to inhibit the inappropriate response, so that the response selection mechanism is somehow corrupted.

Gilbert (2001) suggests that a canonical form of such 'corruption' is the excitation of modes that, in other circumstances, represent normal evolutionary adaptations, a matter to which we will return at some length below.

**6. 'Rational thought'** Although the Cartesian dichotomy between rational thought and emotion may be increasingly suspect, nonetheless humans, like many other animals, do indeed conduct individual rational cognitive decision-making as most of us would commonly understand it. Various forms of dementia involve characteristic patterns of degradation in that ability.

**7. Sociocultural network** Humans, however, are particularly noted for a hypersociality which inevitably enmeshes us all in group processes of decision, i.e. collective cognitive behavior within a social network, tinged by an embedding shared culture. For humans, culture is truly fundamental. Durham (1991) argues that genes and culture are two distinct but interacting systems of inheritance within human populations. Information of both kinds has influence, actual or potential, over behaviors, which creates a real and unambiguous symmetry between genes and phenotypes on the one hand, and culture and phenotypes, on the other. Genes and culture are best represented as two parallel lines or tracks of hereditary influence on phenotypes.

Much of hominid evolution can be characterized as an interweaving of genetic and cultural systems. Genes came to encode for increasing hypersociality, learning, and language skills. The most successful populations displayed increasingly complex structures that better aided in buffering the local environment (e.g. Bonner, 1980).

Successful human populations seem to have a core of tool usage, sophisticated language, oral tradition, mythology, music, magic, medicine, and religion, and decision making skills focused on relatively small family/extended family social network groupings. More complex social structures are built on the periphery of this basic object (e.g. Richerson and Boyd, 1995). The human species' very identity may rest on its unique evolved capacities for social mediation and cultural transmission. These are particularly expressed through the cognitive decision making of small groups facing changing patterns of threat and opportunity, processes in which we are all embedded and all participate.

## Cognition as language

Atlan and Cohen (1998) argue that the essence of cognition is comparison of a perceived external signal with an internal, learned picture of the world, and then, upon that comparison, the choice of one response from a much larger repertoire of possible responses. Clearly, from this perspective, the Thayer and Lane vision of the emotional process is cognitive. Upon recognition of a new perturbation in the surrounding environment, emotional cognition evaluates and choose from several possible responses: no action necessary, or one or a few succinct emotions elicited. Upon appropriate conditioning, the emotional response is able to accelerate the decision process as to what large-scale, general response may be needed, much as the immune system has a more efficient response to second pathogenic challenge once the initial infection has become encoded in immune memory. Similar considerations apply to individual rational thought and to culturally-sculpted group processes.

Following the approach of Wallace (2000, 2002a), we make a very general model of this process.

Pattern recognition-and-response, as we characterize it, proceeds by convoluting an incoming external sensory signal with an internal ongoing activity – the learned picture of the world – and, at some point, triggering an appropriate action based on a decision that the pattern of sensory activity requires a response. We need not specify how the pattern recognition system is trained, and hence we adopt a weak model, regardless of learning paradigm, which can itself be more formally described by the Rate Distortion Theorem. We will, fulfilling Atlan and Cohen's (1998) criterion of meaning-from-response, define a language's contextual meaning entirely in terms of system output.

The model, an extension of that presented in Wallace (2000), is as follows.

A pattern of 'sensory' input, say an ordered sequence $y_0, y_1, ...$, is mixed in a systematic way with internal 'ongoing' activity, the sequence $w_0, w_1, ...$, to create a path of composite signals $x = a_0, a_1, ..., a_n, ...$, where $a_j = f(y_j, w_j)$ for a function $f$. An explicit example will be given below. This path is then fed into a highly nonlinear 'decision oscillator' which generates an output $h(x)$ that is an element of one of two (presumably) disjoint sets $B_0$ and $B_1$. We take

$$B_0 \equiv b_0, ..., b_k,$$

$$B_1 \equiv b_{k+1}, ..., b_m.$$

Thus we permit a graded response, supposing that if

$$h(x) \in B_0$$

the pattern is not recognized, and if

$$h(x) \in B_1$$

the pattern is recognized and some action $b_j, k+1 \leq j \leq m$ takes place.

We are interested in composite paths $x$ which trigger pattern recognition-and-response exactly once. That is, given a fixed initial state $a_0$, such that $h(a_0) \in B_0$, we examine all possible subsequent paths $x$ beginning with $a_0$ and leading exactly once to the event $h(x) \in B_1$. Thus $h(a_0, ..., a_j) \in B_0$ for all $j < m$, but $h(a_0, ..., a_m) \in B_1$.

For each positive integer $n$ let $N(n)$ be the number of paths of length $n$ which begin with some particular $a_0$ having $h(a_0) \in B_0$ and lead to the condition $h(x) \in B_1$. We shall call such paths 'meaningful' and assume $N(n)$ to be considerably less than the number of all possible paths of length $n$ – pattern recognition-and-response is comparatively rare. We further assume that the finite limit

$$H \equiv \lim_{n \to \infty} \frac{\log[N(n)]}{n}$$

both exists and is independent of the path $x$. We will – not surprisingly – call such a cognitive process *ergodic*.

Note that disjoint partition of 'state space' may be possible according to sets of states which can be connected by 'meaningful' paths, leading to a 'natural' coset algebra of the system, a matter of some importance we will not pursue here.

We may thus define an ergodic information source **X** associated with stochastic variates $X_j$ having joint and conditional probabilities $P(a_0, ..., a_n)$ and $P(a_n|a_0, ..., a_{n-1})$ such that appropriate joint and conditional Shannon uncertainties may be defined which satisfy the relations (Cover and Thomas, 1991; Ash, 1990)

$$H[\mathbf{X}] = \lim_{n \to \infty} \frac{\log[N(n)]}{n} =$$

$$\lim_{n \to \infty} H(X_n|X_0, ..., X_{n-1}) =$$

$$\lim_{n \to \infty} \frac{H(X_0, ..., X_n)}{n}.$$

(1)

We say this information source is *dual* to the ergodic cognitive process.

The Shannon-McMillan Theorem provides a kind of 'law of large numbers' and permits definition of the Shannon uncertainties in terms of cross-sectional sums of the form

$$H = -\sum P_k \log[P_k],$$

where the $P_k$ are taken from a probability distribution, so that $\sum P_k = 1$. Again, Cover and Thomas (1991) or Ash (1990) provide algebraic details.

It is important to recognize that different (generalized) languages will be defined by different divisions of the total universe of possible responses into various pairs of sets $B_0$ and $B_1$, or by requiring more than one response in $B_1$ along a path. Like the use of different distortion measures in the Rate Distortion Theorem (e.g. Cover and Thomas, 1991), however, it seems obvious that the underlying dynamics will all be qualitatively similar. Nonetheless, dividing the full set of possible responses into the sets $B_0$ and $B_1$ may itself require 'higher order' cognitive decisions by another module or modules, suggesting the necessity of choice within a more or less broad set of possible languages of thought. This would directly reflect the need to 'shift gears' according to the different challenges faced by the organism. A critical problem then becomes the choice of a normal – zero-mode – language among a very large set of possible languages representing the 'excited states' accessible to the system. This is a fundamental point which we explore below in various ways.

In sum, meaningful paths – creating an inherent grammar and syntax – have been defined entirely in terms of system response, as Atlan and Cohen (1998) propose.

We can apply this formalism to the stochastic neuron in a neural network: A series of inputs $y_i^j, i = 1, ...m$ from $m$ nearby neurons at time $j$ to the neuron of interest is convoluted with 'weights' $w_i^j, i = 1, ..., m$, using an inner product

$$a_j = \mathbf{y}^j \cdot \mathbf{w}^j \equiv \sum_{i=1}^{m} y_i^j w_i^j$$

(2)

in the context of a transfer function $f(\mathbf{y}^j \cdot \mathbf{w}^j)$ such that the probability of the neuron firing and having a discrete output $z^j = 1$ is $P(z^j = 1) = f(\mathbf{y}^j \cdot \mathbf{w}^j)$.

Thus the probability that the neuron does not fire at time j is just $1 - P$. In the usual terminology the $m$ values $y_i^j$ constitute the 'sensory activity' and the $m$ weights $w_i^j$ the 'ongoing activity' at time $j$, with $a_j = \mathbf{y}^j \cdot \mathbf{w}^j$ and the path $x \equiv a_0, a_1, ..., a_n, ....$ A more elaborate example is given in Wallace (2002a).

A little work leads to a standard neural network model in which the network is trained by appropriately varying $\mathbf{w}$ through least squares or other error minimization feedback. This can be shown to replicate rate distortion arguments, as we can use the error definition to define a distortion function which measures the difference between the training pattern $y$ and the network output $\hat{y}$ as a function, for example, of the inverse number of training cycles, $K$. As we will discuss in another context, 'learning plateau' behavior emerges naturally as a phase transition in the mutual information $I(Y, \hat{Y})$ driven by the parameter $K$.

Thus we will eventually parametrize the information source uncertainty of the dual information source to a cognitive pattern recognition-and-response with respect to one or more variates, writing, e.g. $H[\mathbf{K}]$, where $\mathbf{K} \equiv (K_1, ..., K_s)$ represents a vector in a parameter space. Let the vector $\mathbf{K}$ follow some path in time, i.e. trace out a generalized line or surface $\mathbf{K}(t)$. We will, following the argument of Wallace (2002b), assume that the probabilities defining $H$, for the most part, closely track changes in $\mathbf{K}(t)$, so that along a particular 'piece' of a path in parameter space the information source remains as close to memoryless and ergodic as is needed for the mathematics to work. Between pieces we impose phase transition characterized by a renormalization symmetry, in the sense of Wilson (1971). See Binney, et al. (1986) for a more complete discussion.

We will call such an information source adiabatically piecewise memoryless ergodic (APME). The ergodic nature of the information sources is a generalization of the law of large numbers and implies that the long-time averages we will need to calculate can, in fact, be closely approximated by averages across the probability spaces of those sources. This is no small matter.

The reader may have noticed parallels in our development with Fodor's speculations on the language of thought (e.g. Fodor, 1975, 1981, 1987, 1990, 1994, 1998, 1999) in which he proposes a Chomskian linguistically complete background natural language as the basis of mental function in humans. We have taken a weak i.e. asymptotic, approach which does not require linguistic completeness, but only that very long series of outputs may be characterized as the product of a dual information source which is adiabatically, piecewise, memoryless ergodic. This allows importation of phase transition and other methods from statistical physics which provide a natural approach to the interaction of cognitive submodules.

### Interacting information sources: punctuated crosstalk

We suppose that the behavior of a cognitive subsystem can be represented by a sequence of states in time, the path $x \equiv x_0, x_1, ....$ Similarly, we assume an external signal of 'structured psychosocial stress' can also be represented by a path $y \equiv y_0, y_1, ...$. These paths are, however, both very highly structured and, within themselves, are serially correlated and can, in fact, be represented by 'information sources' $\mathbf{X}$ and $\mathbf{Y}$. We assume the cognitive process and external stressors interact, so that these sequences of states are not independent, but are jointly serially correlated. We can, then, define a path of sequential pairs as $z \equiv (x_0, y_0), (x_1, y_1), ...$.

The essential content of the Joint Asymptotic Equipartition Theorem, one of the fundamental limit theorems of 20th Century mathematics, is that the set of joint paths $z$ can be partitioned into a relatively small set of high probability which is termed *jointly typical*, and a much larger set of vanishingly small probability. Further, according to the JAEPT, the *splitting criterion* between high and low probability sets of pairs is the mutual information

$$I(X,Y) = H(X) - H(X|Y) = H(X) + H(Y) - H(X,Y)$$

(3)

where $H(X), H(Y), H(X|Y)$ and $H(X,Y)$ are, respectively, the Shannon uncertainties of $X$ and $Y$, their conditional uncertainty, and their joint uncertainty. See Cover and Thomas (1991) or Ash (1990) for mathematical details. As stated above, the Shannon-McMillan Theorem and its variants permit expression of the various uncertainties in terms of cross sectional sums of terms of the form $-P_k \log[P_k]$ where the $P_k$ are appropriate direct or conditional probabilities. Similar approaches to neural process have been recently adopted by Dimitrov and Miller (2001).

The high probability pairs of paths are, in this formulation, all equiprobable, and if $N(n)$ is the number of jointly typical pairs of length $n$, then, according to the Shannon-McMillan Theorem and its 'joint' variants,

$$I(X,Y) = \lim_{n \to \infty} \frac{\log[N(n)]}{n}.$$

(4)

Generalizing the earlier language-on-a-network models of Wallace and Wallace (1998, 1999), we suppose there is a 'coupling parameter' $P$ representing the degree of linkage between the cognitive system of interest and the structured 'language' of external signals and stressors, and set $K = 1/P$, following the development of those earlier studies. Then we have

$$I[K] = \lim_{n \to \infty} \frac{\log[N(K,n)]}{n}.$$

The essential homology between information theory and statistical mechanics lies in the similarity of this expression with the infinite volume limit of the free energy density. If $Z(K)$ is the statistical mechanics partition function derived from the system's Hamiltonian, then the free energy density is determined by the relation

$$F[K] = \lim_{V \to \infty} \frac{\log[Z(K)]}{V}.$$

(5)

$F$ is the free energy density, $V$ the system volume and $K = 1/T$, where $T$ is the system temperature.

We and others argue at some length (e.g. Wallace and Wallace, 1998, 1999; Wallace, 2000; Rojdestvensky and Cottam, 2000; Feynman, 1996) that this is indeed a systematic mathematical homology which, we contend, permits importation of renormalization symmetry into information theory. Imposition of invariance under renormalization on the mutual information splitting criterion $I(X,Y)$ implies the existence of phase transitions analogous to learning plateaus or punctuated evolutionary equilibria in the relations between cognitive mechanism and external perturbation. An extensive mathematical treatment of these ideas is presented elsewhere (e.g. Wallace, 2000, 2002a,b; Wallace et al., 2003).

Elaborate developments are possible. From a the more limited perspective of the Rate Distortion Theorem, a selective corollary of the Shannon-McMillan Theorem, we can view the onset of a punctuated interaction between the cognitive mechanism and external stressors as the literal writing of distorted image of those stressors upon cognition:

Suppose that two (adiabatically, piecewise memoryless) ergodic information sources $\mathbf{Y}$ and $\mathbf{B}$ begin to interact, to talk to each other, i.e. to influence each other in some way so that it is possible, for example, to look at the output of $\mathbf{B}$ – strings $b$ – and infer something about the behavior of $\mathbf{Y}$ from it – strings $y$. We suppose it possible to define a retranslation from the B-language into the Y-language through a deterministic code book, and call $\hat{\mathbf{Y}}$ the translated information source, as mirrored by $\mathbf{B}$.

Define some distortion measure comparing paths $y$ to paths $\hat{y}$, $d(y, \hat{y})$ (Cover and Thomas, 1991). We invoke the Rate Distortion Theorem's mutual information $I(Y, \hat{Y})$, which is the splitting criterion between high and low probability pairs of paths. Impose, now, a parametization by an inverse coupling strength $K$, and a renormalization symmetry representing the global structure of the system coupling.

Extending the analyses, triplets of sequences, $Y_1, Y_2, Z$, for which one in particular, here $Z$, is the 'embedding context' affecting the other two, can also be divided by a splitting criterion into two sets, having high and low probabilities respectively. The probability of a particular triplet of sequences is then determined by the conditional probabilities

$$P(Y_1 = y^1, Y_2 = y^2, Z = z) = \Pi_{j=1}^n p(y_j^1|z_j)p(y_j^2|z_j)p(z_j).$$

(6)

That is, $Y_1$ and $Y_2$ are, in some measure, driven by their interaction with $Z$.

For large $n$ the number of triplet sequences in the high probability set will be determined by the relation (Cover and Thomas, 1992, p. 387)

$$N(n) \propto \exp[nI(Y_1; Y_2|Z)], \quad (7)$$

where splitting criterion is given by

$$I(Y_1; Y_2|Z) \equiv$$

$$H(Z) + H(Y_1|Z) + H(Y_2|Z) - H(Y_1, Y_2, Z).$$

We can then examine mixed cognitive/adaptive phase transitions analogous to learning plateaus (Wallace, 2002b) in the splitting criterion $I(Y_1, Y_2|Z)$. Note that our results are almost exactly parallel to the Eldredge/Gould model of evolutionary punctuated equilibrium (Eldredge, 1985; Gould, 2002).

We can, for the purposes of this work, extend this model to any number of interacting information sources, $Y_1, Y_2, ..., Y_s$ conditional on an external context $Z$ in terms of a splitting criterion defined by

$$I(Y_1; ...; Y_s|Z) = H(Z) + \sum_{j=1}^{s} H(Y_j|Z) - H(Y_1, ..., Y_s, Z), \quad (8)$$

where the conditional Shannon uncertainties $H(Y_j|Z)$ are determined by the appropriate direct and conditional probabilities.

**The generalized cognitive homunculus and its retina**

Cohen (2000) argues for an 'immunological homunculus' as the immune system's perception of the body as a whole. The particular utility of such a thing, in his view, is that sensing perturbations in a bodily self-image can serve as an early warning sign of pending necessary inflammatory response – expressions of tumorigenesis, acute or chronic infection, parasitization, and the like. Thayer and Lane (2000) argue something analogous for emotional response as a quick internal index of larger patterns of threat or opportunity.

It seems obvious that the collection of interacting cognitive submodules we have explored above must also have a coherent internal self-image of the state of the mind/body and its social relationships. This inferred picture, at the individual level, we term the 'generalized cognitive homunculus', (GCH). We shall use the responses of the GCH to characterize physiological/mental responses to both illness and to medical interventions, including drugs, used to treat that illness. Illness and treatment may come to reflect one another in a hall of mirrors reminiscent of Jerne's idiotypic network proposed for the dynamics of the immune system.

Let us suppose we cannot measure either stress or cognitive submodule function directly, but can determine the concentrations of hormones, neurotransmitters, certain cytokines, and other biomarkers, or else macroscopic behaviors, beliefs, feelings, or other responses associated with the function of cognitive submodules according to some natural time frame inherent to the system. This would typically be the circadian cycle in both men and women, and the hormonal cycle in premenopausal women. Suppose, in the absence of extraordinary meaningful psychosocial stress, we measure a series of $n$ biomarker concentrations, behavioral characteristics, other indices at time $t$ which we represent as an $n$-dimensional vector $X_t$. Suppose we conduct a number of experiments, and create a regression model so that we can, in the absence of perturbation, write, to first order, the markers at time $t+1$ in terms of that at time $t$ using a matrix equation of the form

$$X_{t+1} \approx \mathbf{R} X_t, \quad (9)$$

where $\mathbf{R}$ is the matrix of regression coefficients, and we have normalized to a zero vector of constant terms.

Suppose we write a GCH response to short-term perturbation as

$$X_{t+1} = (\mathbf{R}_0 + \delta \mathbf{R}_{t+1}) X_t,$$

where $\delta \mathbf{R}$ represents variation of the generalized cognitive self-image about the basic state $\mathbf{R}_0$.

We impose a (Jordan block) diagonalization in terms of the matrix of (generally nonorthogonal) eigenvectors $\mathbf{Q}_0$ of some 'zero reference state' $\mathbf{R}_0$, obtaining, for an initial condition which is an eigenvector $Y_t \equiv Y_k$ of $\mathbf{R}_0$,

$$Y_{t+1} = (\mathbf{J}_0 + \delta \mathbf{J}_{t+1}) Y_k = \lambda_k Y_k + \delta Y_{t+1} =$$

$$\lambda_k Y_k + \sum_{j=1}^{n} a_j Y_j, \quad (10)$$

where $\mathbf{J}_0$ is a (block) diagonal matrix as above, $\delta \mathbf{J}_{t+1} \equiv \mathbf{Q}_0 \delta \mathbf{R}_{t+1} \mathbf{Q}_0^{-1}$, and *$\delta Y_{t+1}$ has been expanded in terms of a spectrum of the eigenvectors of* $\mathbf{R}_0$, with

$$|a_j| \ll |\lambda_k|, |a_{j+1}| \ll |a_j|.$$

(11)

The essential point is that, provided $\mathbf{R}_0$ has been properly tuned, so that this condition is true, the first few terms in the spectrum of the plieotropic iteration of the eigenstate will contain almost all of the essential information about the perturbation, i.e. most of the variance. We envision this as similar to the detection of color in the optical retina, where three overlapping non-orthogonal 'eigenmodes' of response suffice to characterize a vast array of color sensations. Here, if a concise spectral expansion is possible, a very small number of (typically nonorthogonal) 'generalized cognitive eigenmodes' permit characterization of a vast range of external perturbations, and rate distortion constraints become very manageable indeed. Thus GCH responses – the spectrum of excited eigenmodes of $\mathbf{R}_0$, provided it is properly tuned – can be a very accurate and precise gauge of environmental perturbation.

The choice of zero reference state $\mathbf{R}_0$, i.e. the 'base state' from which perturbations are measured, is, we claim, a highly nontrivial task, necessitating a specialized apparatus.

This is no small matter. According to current theory, the adapted human mind functions through the action and interaction of distinct mental modules which evolved fairly rapidly to help address special problems of environmental and social selection pressure faced by our Pleistocene ancestors (e.g. Barkow et al., 1992). Here we have postulated the necessity of other physiological and social cognitive modules. As is well known in computer engineering, calculation by specialized submodules – e.g. numeric processor chips – can be a far more efficient means of solving particular well-defined classes of problems than direct computation by a generalized system. We suggest, then, that our generalized cognition has evolved specialized submodules to speed the address of certain commonly recurring challenges. Nunney (1999) has argued that, as a power law of cell count, specialized subsystems are increasingly required to recognize and redress tumorigenesis, mechanisms ranging from molecular error-correcting codes, to programmed cell death, and finally full-blown immune attack.

We argue that identification of the designated normal state of the GCH – generalized cognition's self-image of the body and its social relationships – is difficult, requiring a dedicated cognitive submodule within overall generalized cognition. This is essentially because, for the vast majority of information systems, unlike mechanical systems, there are no restoring springs whose low energy state automatically identifies equilibrium: relatively speaking, all states of the GCH are 'high energy' states. That is, active comparison must be made of the state of the GCH with some stored internal reference picture, and a decision made about whether to reset to zero, which is a cognitive process. We further speculate that the complexity of such a submodule must also follow something like Nunney's power law with animal size, as the overall generalized cognition and its image of the self, become increasingly complicated with rising number of cells and levels of linked cognition.

Failure of that cognitive submodule can result in identification of an excited state of the GCH as normal, triggering the collective patterns of systemic activation which, following the argument of Wallace (2003g), constitute certain comorbid mental and chronic physical disorders. This would result in a relatively small number of characteristic eigenforms of comorbidity, which would typically become more mixed with increasing disorder.

In sum, since such 'zero mode identification' (ZMI) is a (presumed) cognitive submodule of overall generalized cognition, it involves convoluting incoming 'sensory' with 'ongoing' internal memory data in choosing the zero state, i.e. defining $\mathbf{R}_0$. The dual information source defined by this cognitive process can then interact in a punctuated manner with 'external information sources' according to the Rate Distortion and related arguments above. From a RDT perspective, then, those external information sources literally write a distorted image of themselves onto the ZMI, often in a punctuated manner: (relatively) sudden onset of a developmental trajectory to comorbid mental disorders and chronic physical disease.

Different systems of external signals – including but not limited to structured psychosocial stress – will, presumably, write different characteristic images of themselves onto the ZMI cognitive submodule, i.e. trigger different patterns of comorbid mental/physical disorder.

A brief reformulation in abstract terms may be of interest. Recall that the essential characteristic of cognition in our formalism involves a function $h$ which maps a (convolutional) path $x = a_0, a_1, ..., a_n, ...$ onto a member of one of two disjoint sets, $B_0$ or $B_1$. Thus respectively, either (1) $h(x) \in B_0$, implying no action taken, or (2), $h(x) \in B_1$, and some particular response is chosen from a large repertoire of possible responses. We discussed briefly the problem of defining these two disjoint sets, and suggested that some 'higher order cognitive module' might be needed to identify what constituted $B_0$, the set of 'normal' states. Again, this is because there is no low energy mode for information systems: virtually all states are more or less high energy states, and there is no way to identify a ground state using the physicist's favorite variational or other minimization arguments on energy.

Suppose that higher order cognitive module, which we now recognize as a kind of Zero Mode Identification, interacts with an embedding language of structured psychosocial stress (or other systemic perturbation) and, instantiating a Rate Distortion image of that embedding stress, begins to include one or more members of the set $B_1$ into the set $B_0$. Recurrent 'hits' on that aberrant state would be experienced as episodes of highly structured comorbid mind/body pathology.

Empirical tests of this hypothesis, however, quickly lead again into real-world regression models involving the interrelations of measurable biomarkers, beliefs, behaviors, feelings, and so on, requiring formalism much like that used above. The GCH can, then, be viewed as a generic heuristic device typifying such regression approaches.

**Therapeutic efficacy, failure, and adverse reactions**

To reiterate, if $\mathbf{X}$ represents the information source dual to 'zero mode identification' in generalized cognition, and if $\mathbf{Z}$ is the information source characterizing structured psychosocial stress, which constitutes an embedding context, the mutual information between them

$$I(\mathbf{X}; \mathbf{Z}) = H(\mathbf{X}) - H(\mathbf{X}|\mathbf{Z}) \tag{12}$$

serves as a splitting criterion for pairs of linked paths of states.

We suppose it possible to parametize the coupling between these interacting information sources by some inverse coupling parameter, $K$, writing

$$I(\mathbf{X}; \mathbf{Z}) = I[K], \tag{13}$$

with structured psychosocial stress as the embedding context.

Invocation of the mathematical homology between equations (4) and (5) permits imposition of renormalization formalism (Wallace, 2000; Wallace et al., 2003a) resulting in punctuated phase transition depending on $K$.

Socioculturally constructed and structured psychosocial stress, in this model having both (generalized) grammar and syntax, can be viewed as entraining the function of zero mode identification when the coupling with stress exceeds a threshold. More than one threshold appears likely, accounting in a sense for the typically staged nature of environmentally caused disorders. These should result in a synergistic – i.e. co-morbidly excited – mixed affective, rationally cognitive, psychosocial, and inflammatory or other physical excited state of otherwise normal response, and represent the effect of stress on the linked decision processes of various cognitive functions, in particular through the identification of a false 'zero mode' of the GCH. This is a collective, but highly systematic, 'tuning failure' which, in the Rate Distortion sense, represents a literal image of the structure of imposed psychosocial stress written upon the ability of the GCH to characterize a normal condition of excitation, causing a mixed excited state of chronically comorbid mental and physical disorder.

In this model different eigenmodes $Y_k$ of the GCH regression model characterized by the matrix $\mathbf{R}_0$ can be taken to represent the 'shifting-of-gears' between different 'languages' defining the sets $B_0$ and $B_1$. That is, different eigenmodes of the GCH would correspond to different required (and possibly mixed) characteristic systemic responses.

If there is a state (or set of states) $Y_1$ such that $\mathbf{R}_0 Y_1 = Y_1$, then the 'unitary kernel' $Y_1$ corresponds to the condition 'no response required', i.e. the set $B_0$.

Suppose pathology becomes manifest, i.e.

$$\mathbf{R}_0 \to \mathbf{R}_0 + \delta \mathbf{R} \equiv \hat{\mathbf{R}}_0,$$

so that some chronic excited state becomes the new 'unitary kernel', and

$$Y_1 \to \hat{Y}_1 \neq Y_1$$

$$\hat{\mathbf{R}}_0 \hat{Y}_1 = \hat{Y}_1.$$

This could represent, for example, chronic inflammation, autoimmune response, persistent depression/anxiety or HPA axis activation/burnout, and so on.

We wish to induce a sequence of therapeutic counterperturbations $\delta \mathbf{T}_k$ according to the pattern

$$[\hat{\mathbf{R}}_0 + \delta \mathbf{T}_1]\hat{Y}_1 = Y^1,$$

$$\hat{\mathbf{R}}_1 \equiv \hat{\mathbf{R}}_0 + \delta \mathbf{T}_1,$$

$$[\hat{\mathbf{R}}_1 + \delta \mathbf{T}_2]Y^1 = Y^2$$

$$... \tag{14}$$

so that, in some sense,

$$Y^j \to Y_1. \tag{15}$$

That is, the mind/body system, as monitored by the GCH, is driven to its original condition.

We may or may not have $\hat{\mathbf{R}}_0 \to \mathbf{R}_0$. That is, actual cure may not be possible, in which case palliation or control is the therapeutic aim.

The essential point is that the pathological state represented by $\hat{\mathbf{R}}_0$ and the sequence of therapeutic interventions $\delta \mathbf{T}_k, k = 1, 2, ...$ are interactive and reflective, depending on the regression of the set of vectors $Y^j$ to the desired state $Y_1$, much in the same spirit as Jerne's immunological idiotypic hall of mirrors.

The therapeutic problem revolves around minimizing the difference between $Y^k$ and $Y_1$ over the course of treatment: that difference represents the inextricable convolution of 'treatment failure' with 'adverse reactions' to the course of treatment itself, and 'failure of compliance' attributed through social construction by provider to patient, i.e. failure of the therapeutic alliance.

It should be obvious that the treatment sequence $\delta\mathbf{T}_k$ represents a cognitive path of interventions which has, in turn, a dual information source in the sense we have previously invoked.

Treatment may, then, interact in the usual Rate Distortion manner with patterns of structured psychosocial stress which are, themselves, signals from an embedding information source. Thus treatment failure, adverse reactions, and patient noncompliance will, of necessity, embody a distorted image of structured psychosocial stress.

In sum, characteristic patterns of treatment failure, adverse reactions, and patient noncompliance reflecting collapse of the therapeutic alliance, will occur in virtually all therapeutic interventions according to the manner in which structured psychosocial stress is expressed as an image within the treatment process. This would most likely occur in a highly punctuated manner, depending in a quantitative way on the degree of coupling of the three-fold system of affected individual, patient/provider interaction, and treatment mode, with that stress.

Given that the principal environment of humans is defined by interaction with other humans and with socioeconomic institutions, these are likely to be very strong effects indeed.

We provide a non-pharmaceutical case history.

## Malaria and the Fulani: non-pharmaceutical medical interventions and structured psychosocial stress

Modiano et al. (1996, 1998, 2001a, b) conducted comparative surveys on three roughly co-resident West African ethnic groups exposed to the same strains of malaria. The Fulani, Mossi, and Rimaibe live in the same conditions of hyperendemic transmission in a Sudan savanna area northeast of Ouagadougou, Burkina Faso. The Mossi and Rimaibe are Sudanese Negroid populations with a long tradition of sedentary farming, while the Fulani are nomadic pastoralists, partly settled and characterized by non-Negroid features of possible Caucasoid origin.

Parasitological, clinical, and immunological investigations showed consistent interethnic differences in *P. Falciparum* infection rates, malaria morbidity, and prevalence and levels of antibodies to various *P. Falciparum* antigens. The data point to a remarkably similar response to malaria in the Mossi and Rimaibe, while the Fulani are clearly less parasitized, less affected by the disease, and more responsive to all antigens tested. No difference in the use of malaria protective measures was demonstrated that could account for these findings. Known genetic factors of resistance to malaria showed markedly *lower* frequencies in the Fulani (Modiano et al, 2001a, b). The differences in the immune response were not explained by the entomological observations, which indicated substantially uniform exposure to infective bites.

In their first study, Modiano et al. (1996) concluded that sociocultural factors are not involved in this disparity, and that the available data support the existence of unknown genetic factors, possibly related to humoral immune responses, determining interethnic differences in the susceptibility to malaria.

In spite of later finding the Fulani in their study region have significantly *reduced* frequencies of the classic malaria-resistance genes compared to the other ethnic groups, Modiano et al. (2001a, b) again concluded that their evidence supports the existence in the Fulani of unknown genetic factor(s) of resistance to malaria.

This vision of strict genetic causality carries consequences, seriously constraining interpretation of the efficacy of interventions. Modiano et al. (1998) report results of an experiment in their Burkina Faso study zone involving the distribution of permethrin-impregnated curtains (PIC) to the three co-resident populations, with markedly different results:

> " The PIC were distributed in June 1996 and their impact on malaria infection was evaluated in [the three] groups whose baseline levels of immunity to malaria differed because of their age and ethnic group. Age- and ethnic-dependent efficacy of the PIC was observed. Among Mossi and Rimaibe, the impact (parasite rate reduction after PIC installation with respect to the pre-intervention surveys) was 18.8 % and 18.5 %, respectively. A more than two-fold general impact (42.8 %) was recorded in the Fulani. The impact of the intervention on infection rates appears positively correlated with the levels of anti-malaria immunity..."

Modiano et al. (1998) conclude from this experiment that the expected complementary role of a hypothetical vaccine is presaged by these results, which also, in their view, emphasize the importance of the genetic background of the population in the evaluation and application of malaria control strategies.

While we fully agree with the importance of their results for a hypothetical vaccine, much in the spirit of Lewontin (2000), we beg to differ with the ad hoc presumptions of genetic causality, which paper over alternatives involving environment and development consistent with these observations.

Recently a medical anthropologist, Andrew Gordon (2000), published a detailed study of Fulani cultural identity and illness:

> "Cultural identity – who the Fulani think they are – informs thinking on illnesses they suffer. Conversely, illness, so very prevalent in sub-Saharan Africa, provides Fulani with a consistent reminder of their distinctive condition... How they approach being ill also tells Fulani about themselves. The manner in which Fulani think they are sick expresses their sense of difference from other ethnic groups. Schemas of [individual] illness and of collective identity draw deeply from the same well and web of thoughts... As individuals disclose or conceal illness, as they discuss illness and the problem of others, they reflect standards of Fulani life – being strong of character not necessarily of body, being disciplined, rigorously Moslem, and leaders among lessors... to be in step with others and with cultural norms is to have pride in the self and the foundations of Fulani life."

The Fulani carried the Islamic invasion of Africa into the sub-Sahara, enslaving and deculturing a number of ethnic groups, and replacing the native languages with their own. This is much the way African Americans were enslaved, decultured, and taught English.

As Gordon puts it,

> " 'True Fulani' see themselves as distinguished by their aristocratic descent, religious commitments, and personal qualities that clearly differ from lowland cultivators. Those in the lowland are, historically, Fulani subjects who came to act like and speak Fulani, but they are thought to be without the right genealogical descent. The separation between pastoralists and agriculturists repeats itself in settlements across Africa. The terms vary from place to place in Guinea, the terms are Fulbhe for the nobles and the agriculturalist Bhalebhe or Maatyubhee; in Burkina Faso, Fulbhe and the agricultural Rimaybhe; and in Nigeria, the Red Fulani and the agricultural Black Fulani... The schemas for the Fulani body describe the differences between them and others. These are differences that justify pride in being Fulani and not Bhalebhe, Maatyubhe, Rimaybhe, or Black Fulani. In Guinea, the word 'Bhalebhe' means 'the black one'. The term 'Bhalebhe' carries the same meaning as 'Negro' did for Africans brought to North America. It effaces any tribal identity...
>
> The control a Fulani exercises over the body is an essential feature of 'the Fulani way.' Being out of control is shameful and not at all Fulani-like... To act without restraint is to be what is traditionally thought of as Bhalebhe...
>
> Being afflicted with malaria – and handling it well – is a significant proof of ethnicity. How Fulani handle malaria may be telling. What they lack in physical resistance to disease they make up in persistence. Though sickly, Fulani men only reluctantly give into malaria and forgo work. To give into physical discomfort is not *dimo*. When malaria is severe for a man he is likely not to succumb to bed, but instead to sit outside of his home socializing."

The contrasting Occam's Razor hypothesis to genetic determinism, then, is that the observed significant difference in both malarial parasitization and the efficacy of non-pharmaceutical intervention between the historically-dominant Fulani and co-resident historically-subordinate ethnic groups in the Ouagadougou region of Burkina Faso is largely accounted for by longitudinal and cross-sectional factors of structured psychosocial stress, synergistically intersecting with medical intervention, particularly in view of the *lower* frequencies of classic malaria-resistance genes found in the Fulani.

It is not that the Fulani aren't parasitized, or that the 'Fulani way' prevents disease, but that the population-level burdens of environment are modulated by historical development, and these are profoundly different for former masters and former slaves.

### Conclusion: The coming plagues of drug failure, noncompliance, and adverse reactions in the US

Structured psychosocial stress can, from our development, write an image of itself onto the success or failure of individual-level therapeutic intervention, drug-related or not.

The punctuated nature of individual-level response to the coupling with structured stress should, when averaged across a population, reflect itself in a nonlinear 'dose-response' relation between environmental indices of stress and indices of physiological impact, much as in figure 1. Apartheid systems, which structure stress in 'Western' societies, following the models of Fanon (1966), Memmi, (1967, 1969), or Wilkinson (1996), are generally seen as frozen, Manichean, structures entangling dominant and subordinate populations in a synergistically dehumanizing pathogenic relationship adversely affecting both, although the powerful are, as always, relatively healthier than those they dominate.

Within the United States, however, the apartheid system is changing in such a way as to enmesh increasingly larger populations in its outfalls. Our own studies show how the post WW-II 'hollowing out' of major US urban centers through contagious urban decay and other forms of policy-driven 'urban desertification' has pumped social disintegration – carrying with it both contagious disease and behavioral pathology – into the commuting field surrounding central cities. These phenomena, in turn, have likewise diffused down along the US urban hierarchy itself, from larger to smaller metropolitan regions (e.g. D. Wallace and R. Wallace, 1998; R. Wallace and D. Wallace, 1997; D. Wallace, 2001). R. Wallace et al. (1997) explicitly demonstrate, for eight large US urban centers, how spreading social disintegration entrained AIDS, tuberculosis, and violent crime into surrounding suburban counties, the most 'hollowed out' central cities being the most pathogenic.

At present, chronic disease epidemics of asthma, obesity, and diabetes, are spreading across the US, even marching up the social hierarchy. It is our contention that these represent the enmeshment of more and more subpopulations into relations of pathogenic social hierarchy like figure 1 (e.g. Wallace et al., 2003c, d).

Figures 2, 3, and 4 display data related to that argument.

We predict that, not only will chronic developmental disorders continue to diffuse socially and geographically, but failures of drug efficacy and patient compliance, and adverse drug reactions, which in a literal sense mirror the same enmeshing patterns of structured psychosocial stress creating the disorders, will likewise diffuse between and within both social subgroups and geographic regions in a similar manner. Current ideologies of genetic determinism however, have created significant impediments to empirical studies of this hypothesis. These ideological constraints on science may, in the long run, prove very expensive indeed to the pharmaceutical industry. The example of Lysenkoism in the agriculture of the Soviet Union may well be mirrored by the effects of ideological geneticism in Western medicine.

In sum, slavery, racism, apartheid, ethnic cleansing, and their collective aftermath, in concert with draconian economic inequality compounded by deindustrialization in the aftermath of the Cold War, form the context for patterns of both disease onset and progression, and medical treatment efficacy affecting ethnic minorities in the United States and other subject populations elsewhere in the world. This is a context which is not containable within those populations, but enmeshes dominant subgroups as well (e.g. Memmi, 1967, 1969), as exemplified by figures 2-4.

The result of Exner et al. (2001) regarding a more than three-fold difference in heart failure drug efficacy between white and African-American males in the US does not need invocation of genetic differences for its explanation. Following Lewontin's (2000) lead, we invoke instead the triple helix of genes, environment and development, in both disease and in response to, and compliance with, medical intervention against disease. Given that genetic differences between chimpanzees are generally far greater than genetic differences between humans, the Occam's Razor explanation for differences in response to medical intervention would seem to lie primarily in matters of environment and development.

Independent of the mathematical modeling exercise which has led us to this conclusion – a conclusion others have reached without recourse to mathematics, (e.g. Braun, 2002) – this is an empirically testable hypothesis. The most direct test would be the reevaluation of existing drug trial data from the geographic perspective of 'neighborhood effects', since, within the US, as a result of American Apartheid, 'race is place'.

While most drug trials are stratified by the usual population divisions of age, sex, race/ethnicity, income/education, and so on, they are seldom if ever examined from a formally geographic perspective. Detection of neighborhood effects in the spectrum of drug efficacy, compliance, and adverse reactions, would be a considerable advance. Changes in the pattern of such neighborhood effects over time, a likely consequence of diffusing social disintegration in the US, would serve to monitor the sociogeographic spread of drug failure and decay of the therapeutic alliance. This is likely to follow the classic mode of hierarchical hopscotch from major metropolitan regions like New York and Los Angeles, to smaller ones, then from central city to suburbs, and finally along social networks (e.g. Wallace and Wallace, 1997).

A second, and related line of work, would see development of a short preclinical interview instrument exploring (1) patterns of cross-sectional psychosocial stress and (2) geographic history of residence, and testing whether such an instrument would account for significant variance in drug efficacy, compliance, and adverse reactions. While this would certainly not substitute for programs of reform to ameliorate diffusing patterns of pathogenic social hierarchy and social disintegration, it might nonetheless prove of some individual-level value in adjusting drug regimens.

A third extension would involve expansion of this analysis to medical intervention beyond simple drug treatments, for example surgical outcomes and the results of psychosocial interventions. These too should show neighborhood effects indicating spatially and socially diffusing structured psychosocial stress according to the classic pattern.

A simple interview protocol would likely account for at least as much variance in prediction of treatment efficacy, compliance, and adverse reactions as more expensive and difficult genetic testing. It is even (remotely) possible that the two techniques could be used in parallel to provide a reasonably full characterization of the version of Lewontin's 'triple helix' which defines therapeutic response, a significant step toward a truly biological medicine far more in concert with the realities of human hypersociality than current simplistic and ideologically driven genetic determinism.

On a more somber note, however, while some individual-level tuning of conventional medical treatment does seem possible, it is painfully obvious that no medical system can ameliorate significant population-level health disparities in the absence of aggressive affirmative action to redress both the persisting burdens of history and current policies driving pathogenic hierarchy and social disintegration. This being said, it seems obvious that the economic well-being of the drug industry, if not precisely the health of the populations it services, will depend on a fuller understanding of the sociogeographic diffusion of drug failure, compliance failure, and adverse reactions. Pharmacogenetic theories of race constitute a significant barrier to such understanding.

## Acknowledgments

This work benefited from support under NIEHS Grant I-P50-ES09600-05.

**Figure Captions**

**Figure 1**. Redisplay of data from Singh-Manoux et al. (2003). Sex-specific dose-response curves of age-adjusted prevalence of self-reported ill-health vs. self-reported status rank, Whitehall II cohort, 1997 and 1999. 1 is high and 10 is low status. The curves for male and female are virtually identical, and the upper point is very near the 'EC-50' level in this population i.e. the 'effective concentration' at which fifty percent of subjects show physiological response. Self-reported health is a highly significant predictor of later morbidity and mortality.

**Figure 2**. US Black vs. White diabetes death rates (per 100,000), 1979-97. While the Black rates are uniformly higher than the White, the coupling between them is very strong indeed, suggesting that, in the words of one researcher, "concentration is not containment" for chronic as well as for infectious diseases like AIDS and tuberculosis.

**Figure 3**. Same as figure 2 for US hypertension death rates.

**Figure 4**. Percent of total US income accounted for by the highest five percent of the population vs. integral of manufacturing job loss since 1980. Manufacturing job loss represents the permanent, cumulative, dispersal of social, political, and economic capital for large sectors of the US population.

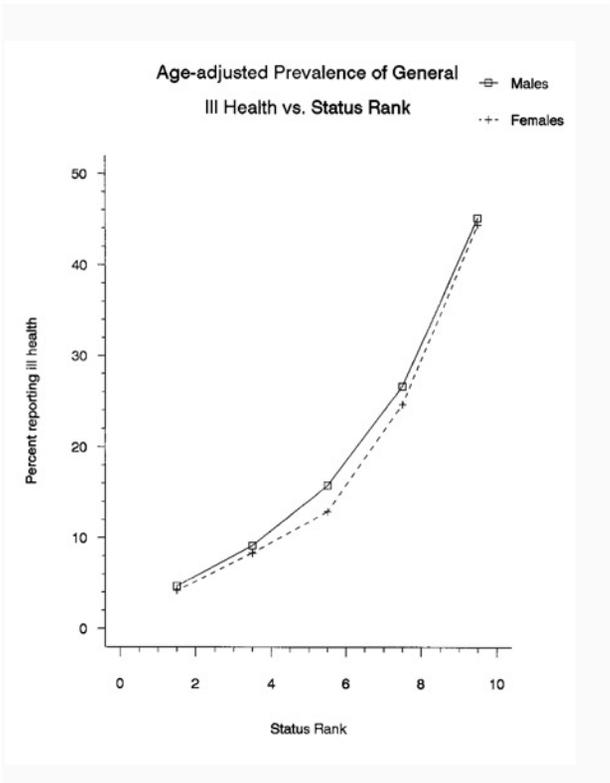
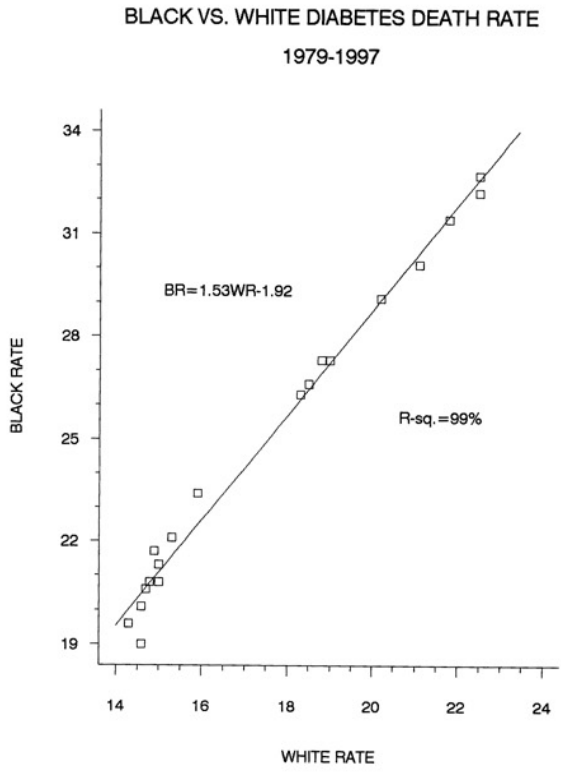
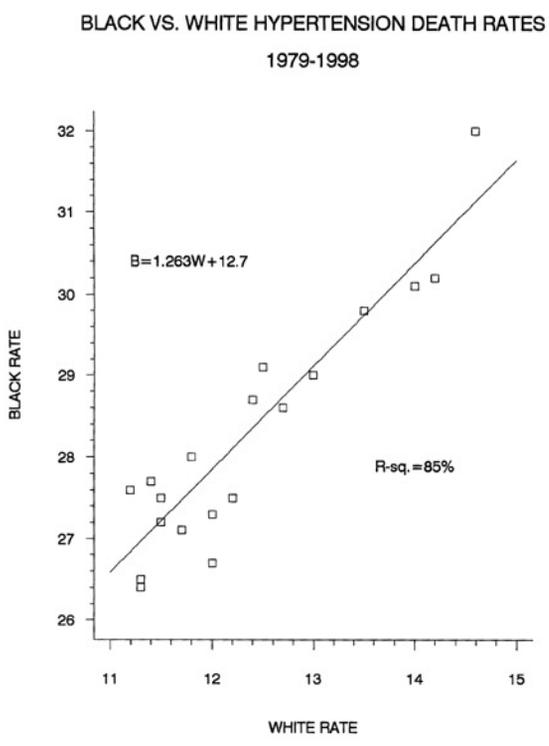
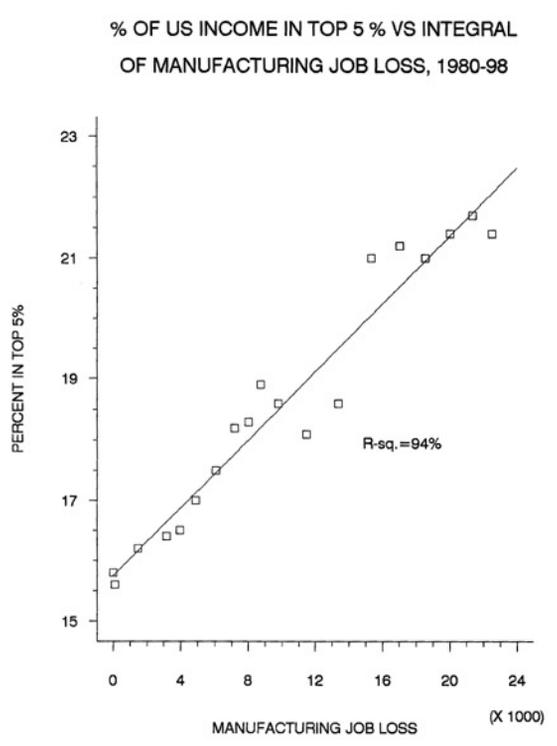